\documentclass[aps,twocolumn,prl,10pt,showpacs,amsmath,amssymb]{revtex4}

\usepackage{slashed}
\usepackage[dvips]{color}
\usepackage[dvips]{epsfig}
\usepackage{latexsym}
\usepackage{bm}
\usepackage{upgreek}
\usepackage{mathrsfs}
\usepackage{times}
\usepackage{amsthm}
\usepackage{amssymb}
\usepackage{epsfig}
\usepackage{graphicx}
\usepackage{amsmath}

\begin{document}

\title{On thermal Nieh-Yan anomaly in topological Weyl materials}

\author{J. Nissinen}

\affiliation{Low Temperature Laboratory, Department of Applied Physics, Aalto University, P.O. Box 15100, FI-00076
Aalto, Finland}

\email{jaakko.nissinen@aalto.fi}

\author{G.E. Volovik}

\affiliation{Low Temperature Laboratory, Department of Applied Physics, Aalto University, P.O. Box 15100, FI-00076
Aalto, Finland}
\affiliation{Landau Institute for Theoretical Physics RAS, Kosygina 2,
119334 Moscow, Russia}

\email{volovik@boojum.hut.fi}

\date{\today}

\begin{abstract}
We discuss the possibility of a gravitional Nieh-Yan anomaly of the type $\partial_\mu j^\mu_5 =\gamma T^2{\cal T}^a\wedge {\cal T}_a$ in topological Weyl materials, where 
 $T$ is temperature and ${\cal T}^a$ is the effective or emergent torsion. As distinct from the non-universal parameter $\Lambda$ in the conventional (zero temperature) Nieh-Yan anomaly --- with canonical dimensions of momentum --- the parameter $\gamma$ is dimensionless. This suggests that the dimensionless parameter is fundamental, being determined by the geometry, topology and number of the number of chiral quantum fields without any explicit non-universal UV scales. This conforms with previous results in the literature,  as well as spectral flow calculations using torsional magnetic field at finite temperature.
\end{abstract}

\maketitle

\section{Introduction}

In non-relativistic topological matter, quasi-relativistic description of low-energy quasiparticles with linear spectrum phenomena may emerge \cite{Volovik2003, Horava05}. In particular, in three spatial dimensions at a generic (two-fold) degenerate fermion band crossing at momentum $\mathbf{p}_W$, the Hamiltonian is of the Weyl form \cite{Herring1937, Abrikosov1971, NielsenNinomiya83, Volovik2003}
\begin{align}
H_{W} = \sigma^a e^{i}_a (p- p_W)_i + \cdots \label{eq:WeylHamiltonian}
\end{align}
where the $e^{i}_a = \partial_{p_i} H(\mathbf{p})\vert_{p_W}$ are the linear coefficients of the Hermitean Pauli matrices $\sigma^a$, close to the Weyl node at $\mathbf{p}_W$. The net chirality $\sum_{\mathbf{p}_W} \textrm{sgn}(\det e^{i}_a)$ vanishes. For slowly varying parameters in the Hamiltonian operator $i\partial_t -H_W$, the semi-classical fields $e^{\mu}_a(x) = \{\delta^0_{a}, e^i_a\}$ are promoted to background spacetime tetrad fields with dimensions of unity for temporal indices $e_a^0$ and velocity for spatial $e^i_a$. The shift of the Weyl node $\mathbf{p}_W$ acts as an emergent (axial) gauge field with emergent Lorentz symmetry to the linear order. If the fermions are charged, the fermions can couple to the electromagnetic vector potential via minimal coupling. These background fields imply the chiral anomaly for the low-energy massless quasiparticles. For the applications of the chiral anomaly in Weyl semimetal and Weyl superfluids/superconductors, see e.g. \cite{NielsenNinomiya83, Volovik1986a, Volovik2003, ZyuzinBurkov12}.

In particular, the non-trivial coordinate dependence (torsion) related to the tetrads $e_a^{\mu}(x)$ in \eqref{eq:WeylHamiltonian} can lead to the gravitational Nieh-Yan anomaly \cite{NiehYan1982, Yajima96, ChandiaZanelli97, Nissinen2019}. Here we will discuss this in the presence of finite temperature \cite{NissinenVolovik2019, Stone2019}.

\section{Torsional anomaly}

For spacetimes with torsion (and curvature), Nieh and Yan \cite{NiehYan1982} introduced the 4-dimensional invariant
\begin{equation}
N=\mathcal{T}^a \wedge \mathcal{T}_a+ e^a \wedge e^b\wedge R_{ab} \,
\label{N}  
\end{equation}
where $e^a_{\mu} dx^{\mu}$ is the local tetrad field and $\mathcal{T}^a = de^a + \omega^a_{\ b} \wedge e^b$ and $R^{ab}=d\omega^a_{b} + \omega^a_{\ c} \wedge \omega^c_{\ b}$, in terms of the tetrad and spin-connection $\omega^a_{\ b} = \omega^a_{\mu b} dx^{\mu}$. This invariant can be written as
\begin{equation}
N=dQ \,\,,  \,\, Q=e^a\wedge {\cal T}_a \,,
\label{N2}  
\end{equation}
using the associated Bianchi identities, which says that $N$ is a locally exact 4-from independent from $\textrm{tr} (R \wedge R)$ and dual of the scalar curvature $\mathcal{R}$ in the presence of non-zero torsion. It can be associated with a difference of two topological terms, albeit in terms of an embedding to five dimensions \cite{ChandiaZanelli97}. In terms of four-dimensional chiral fermions on such a spacetime, it has been suggested that this invariant contributes to the axial anomaly, i.e. the anomalous production of the chiral current:
\begin{equation}
\partial_\mu j_5^\mu =  \frac{\Lambda^2}{4\pi^2} N({\bf r},t) \,,
\end{equation}
where $j^{\mu}_5$ is the axial current (tensor)density and the non-universal parameter $\Lambda$ has dimension of relativistic momentum (mass) $[\Lambda]=[1/L] = [M]$ and is determined by some ultraviolet energy scale.

There has been several attempts to consider the Nieh-Yan anomaly in condensed matter systems with Weyl fermions like \eqref{eq:WeylHamiltonian}, see e.g. \cite{Parrikar2014, SunWan2014, FerreirosEtAl19, Nissinen2019}. However, in non-relativistic systems the relativistic high-energy cut-off $\Lambda$ is not a well defined parameter and it can be anisotropic. The complete UV theory is non-Lorentz invariant of course and the linear, quasirelativistic Weyl regime is valid at much lower scales. Moreover, the anomalous hydrodynamics of superfluid $^3$He at zero temperature suggests that the chiral anomaly is completely exhausted by the emergent axial gauge field corresponding to the shift of the node or, conversely, the NY anomaly term for local Lorentz invariance along the uniaxial symmetry direction, leaving no space for the chiral gravitational Nieh-Yan and gauge anomalies to coexist. Nevertheless, it was shown in Ref. \cite{Nissinen2019} that the low-energy theory satisfies the symmetries and conservations laws related to an emergent quasirelativistic spacetime with torsion and $\Lambda$ is determined from the UV-scale where the linear Weyl approximation breaks down, as dictated by the underlying $p$-wave BCS Fermi superfluid.

In terms of effective low-energy effective actions, it seems that the fully relativistic analogs work unambiguously only for terms in the effective action with dimensionless coefficients. Perhaps the most well-known example being the 2+1-dimensional topological Chern-Simons (CS) terms describing the Quantum Hall effect. Gravitational Chern-Simons terms similarly are quantized in terms of chiral central charge which has relation to thermal transport and the boundary conformal field theory \cite{Volovik90, ReadGreen01}. The CS action was recently generalized to crystalline topological insulators in higher (three) odd space dimensions. The latter higher dimensional Chern-Simons terms are expressed via of elasticity tetrads $E$ with dimension $[E]=[1/L]=[M]$, which allows to have topological terms of the type $E\wedge A\wedge d A$ with quantized dimensionless coefficients \cite{NissinenVolovik2018b, NissinenVolovik2018, SongEtAl2019}. 

Another such example is the temperature correction to curvature effects, with $\delta S_{\rm eff}$ = $\int T^2{\cal R}$ in the low-energy action \cite{VolovikZelnikov2003}. This represent the analog of the gravitational coupling (Newton constant) in the low-energy action where the curvature scalar ${\cal R}$ is some analog of scalar spacetime curvature. Since $[T]^2[{\cal R}]=[M]^4$, the coefficient of this term in the low-energy theory is dimensionless, and thus can be given in terms of universal constants. Actually, the coefficient of this term is background independent, i.e. it is fully determined by the number of the  fermionic and bosonic species in the effective theory on flat background, and thus works both in relativistic and non-relativistic systems \cite{VolovikZelnikov2003}. Nevertheless, in general condensed matter systems this term is typically subdominant. The more relevant zero-temperature terms, such as $\int G^{-1} {\cal R}$ in the Einstein action, always contain 
the dimensionful parameters, such as $[G]=[M]^{-2}$ and are not reproduced in realistic condensed matter systems \cite{Volovik2003}. 

The same universal behavior takes place with the terms describing the chiral magnetic and chiral vortical effects in Weyl superfluid $^3$He-A, where the coefficients are dimensionless \cite{VolovikVilenkin2000,Volovik2003}. Similarly, the coefficient of the $R\wedge R$ graviational anomaly in chiral Weyl systems affects the thermal transport coefficients in flat space \cite{Lucas2016, GoothEtAl17}. These coefficients are fundamental, being determined by the underlying degrees of freedom in addition to symmetry, topology and geometry.

\section{Temperature correction to the Nieh-Yan term} 

It is quite possible that the same situation may hold for the temperature correction to the Nieh-Yan anomaly term \cite{NiehYan1982}. The zero-temperature anomaly term in the axial current production  $\Lambda^2({\cal T}^a \wedge {\cal T}_a+ e^a\wedge e^b\wedge R_{ab})$ is still not confirmed in general. On one hand the ultra-violet cut-off parameter $\Lambda$ is not well-defined in relativistic field theory with fundamental chiral fermions. On the other hand, such a cut-off is not in general available in non-relativistic matter with quasi-relativistic low-energy chiral fermions and can be anisotropic \cite{Nissinen2019} or even zero. However, the term of the form $T^2({\cal T}^a\wedge {\cal T}_a+ e^a\wedge e^b\wedge R_{ab})$ has the proper dimensionality $[M]^{4}$, and its prefactor could be a universal constant, being expressed via some invariant related to the degrees of freedom.

For concreteness, we focus on the finite temperature Nieh-Yan anomaly in chiral $p$-wave Weyl superfluid (such as $^3$He-A) with
\begin{equation}
\partial_\mu j_5^\mu = \gamma   T^2 N({\bf r},t) \,,
\label{T2NiehYan}
\end{equation}
and check whether the dimensionless parameter $\gamma$ can be universal. We now use the result obtained by Khaidukov and Zubkov \cite{Zubkov2018} and Imaki and Yamamoto \cite{Imaki2019} for the finite temperature contribution to the chiral current. For single chiral (complex) fermion, one has for the chiral current
\begin{equation}
j^k_5= -  \frac{T^2}{24}   \epsilon^{0kij}T^0_{ij} \,.
\label{j5}
\end{equation}
We assume that this current can be covariantly generalized  to the 4-current:
\begin{equation}
j^\mu_5= -  \frac{T^2}{24}  \epsilon^{\mu\nu\alpha\beta} e_{\nu a}T^a_{\alpha\beta} \,.
\label{j5general}
\end{equation}
Then one obtains the divergence
\begin{equation}
\partial_{\mu} j^\mu_5= -  \frac{T^2}{48}  \epsilon^{\mu\nu\alpha\beta} T_{a\mu \nu}T^a_{\alpha\beta}
 \,.
\label{j5nonconservation}
\end{equation}
In the presence of curvature $R(\omega)$, this becomes the temperature correction to the full Nieh-Yan term in Eq. (\ref{T2NiehYan}), where now the non-universal cut-off $\Lambda$ is substituted by the well defined temperature $T$, and the dimensionless parameter $\gamma=1/12$ \cite{NissinenVolovik2019}:
\begin{equation}
\partial_\mu j^\mu_5=-  \frac{T^2}{12}   N({\bf r},t) \,.
\label{j5nonconservationR}
\end{equation}
This result has been confirmed in Ref. \cite{Stone2019} by a direct calculation of the spectral flow of Landau levels in the presence of a constant torsional magnetic field $T^3_{\mu\nu}$ at finite temperature. Note that the local relativistic (Tolman) temperature $T=T_0/\vert e^0_t \vert$ enters the local anomaly, while the constant $T_0$ is the global equilibrium temperature of the condensed matter system \cite{VolovikZelnikov2003}. In \eqref{eq:WeylHamiltonian} we have $e^0_t = -1$. 

\section{From relativistic physics to chiral Weyl superfluid} 

Let us apply this equation to the chiral superfluid $^3$He-A. The chiral anomaly for the Weyl quasiparticles leads to the anomalous production of the linear momentum \cite{Volovik2003}. 
The reason for that is that the spectral flow of chiral quasiparticles is accompanied by the spectral flow of finite linear momentum ${\bf p}_W$ at the Weyl point. For a single chiral fermion, this is proportional to $p_Fj^{0}$ ,which is not conserved in the presence of non-trivial background fields. In $^3$He-A  there are two spin-degenerate Majorana-Weyl points with opposite chirality and opposite momenta, ${\bf p}_{\pm W}=\pm p_F{\hat{\bf l}}$, where $\hat{\bf l}$ is the anisotropy axis due to the orbital 
momentum of the liquid. Then the anomalous production of the linear momentum densities from the two Weyl points sum up to $j^0_5$and one has:
\begin{equation}
\dot {\bf P}_{\rm anom} = -p_F \hat{\bf l}\, (\partial_t j^0_5)\,.
\label{anomalousProductionP}
\end{equation}
This is consistent with the anomalous superfluid hydrodynamics: total momentum in the system is conserved. Thus Eq.(\ref{j5nonconservation}) gives the temperature correction to this
anomalous momentum production:
\begin{eqnarray}
\dot {\bf P}_{\rm anom}(T)-\dot {\bf P}_{\rm anom}(0) =
\nonumber
p_F \hat{\bf l}\,\partial_t (j^0_5\vert_T - j^{0}_5\vert_{T=0})   \\
= p_F \hat{\bf l} \,\frac{T^2}{24} N = p_F \hat{\bf l} \,\frac{T^2}{48} \epsilon^{\mu\nu\alpha\beta} T_{a\mu \nu}T^a_{\alpha\beta} \,.
\label{dotP}
\end{eqnarray}

Let us express Eq.(\ref{dotP}) in terms of the hydrodynamic variables of chiral superfluid.
Here we ignore the superfluid velocity. Then the only hydrodynamic broken symmetry variable is the unit vector of the orbital 
momentum  $e_3^i = \hat{\bf l}$.   The full (inverse) vierbein $e^{\mu}_a$ in \eqref{eq:WeylHamiltonian} in the vicinity of the Weyl point have the form:
\begin{eqnarray}
{e}^{\mu}_a &=&
\,\left(\begin{array}{cc}1 & 0\\
 0 & c_\bot \hat{\bf m} \\ 0 & c_\bot \hat{\bf n} \\ 0 &c_\parallel {\bf l}\end{array} \right)  , \quad a,\mu = 0,1,2,3  \,.
\end{eqnarray}
Here $c_\bot(\hat{\bf m}+i\hat{\bf n}) = \frac{\Delta_0}{p_F}(\hat{\bf m}+i\hat{\bf n})$ is the order parameter in the $p+ip$ chiral superfluid;  
$\hat{\bf l}=\hat{\bf m}\times\hat{\bf n}$ is the unit vector in the direction of the orbital angular momentum of Cooper pairs; $c_\parallel = p_F/m$ and $c_\bot =\Delta_0/p_F$ are anisotropic effective speeds of light in Weyl equation \eqref{eq:WeylHamiltonian} normal to the quasiparticle Fermi-surface $\hat{\bf l}$ and the transverse directions, respectively. In the weak coupling BCS theory $c_\bot \ll c_\parallel$.
For the vierbein ${e}^i_a {e}^a_j = \delta^i_j$ we have
\begin{eqnarray}
e^a_{\mu} &=&
\,\left(\begin{array}{cccc}1 & 0 & 0&0 \\
 0 & \frac{1}{c_\bot}\hat{\bf m}  & \frac{1}{c_\bot}\hat{\bf n} & \frac{1}{c_\parallel}\hat{\bf l}\end{array} \right)  , \quad a,\mu = 0,1,2,3 \label{Eu}
\end{eqnarray}

The torsion (and curvature) tensors corresponding to this tetrad were calculated in Ref. \cite{Nissinen2019}, producing the following Nieh-Yan invariant $N$, if the effect of the superfluid velocity (curvature) is ignored:
\begin{equation}
N = \frac{2}{c_\bot^2}  (\hat{\bf l} \cdot (\partial_t  \hat{\bf l} \times (\hat{\bf l} \cdot \nabla)\hat{\bf l}))
 \,.
\label{NNissinen}
\end{equation}
Assuming that Eq.(\ref{dotP}) is applicable to $^3$He-A, one obtains that the temperature correction to momentum production determined by the Nieh-Yan invariant has the following form:
\begin{equation}
\dot {\bf P}_{\rm anom}(T) =  \dot {\bf P}_{\rm anom}(0) 
 +  p_F \hat{\bf l} \,\frac{T^2}{24c_\bot^2} (\partial_t  \hat{\bf l} \cdot (\nabla\times\hat{\bf l}))
 \,.
\label{dotP2}
\end{equation}

\section{Conclusions}

We suggested that the temperature correction to the gravitational Nieh-Yan anomaly can be universal for chiral Weyl fermions. It is fully determined by the quasirelativistic physics in the vicinity of the Weyl node, and does not depend on the non-universal cut-off as distinct from the zero temperature term. We found the universal prefactor using the relativistic regularization scheme of Refs. \cite{Zubkov2018,Imaki2019, NissinenVolovik2019}. More recently this prefactor has been confirmed in Ref. \cite{Stone2019} in a finite temperature spectral flow calculation due to torsional magnetic fields \cite{Volovik85, Parrikar2014}. 

We compared the result to chiral $p$-wave superfluid with Weyl points, where it is known that the hydrodynamics of $^3$He-A experiences the chiral anomaly due to non-trivial order parameter textures and superflow. The simplest way to compute this is the spectral flow through the Weyl points \cite{Volovik85, Volovik2003}. The spectral flow of momentum depends only on the density of states at the node. Therefore the calculation using the relativistic invariant physics near the gap node gives the same result as the full BCS Fermi-liquid using the semi-classical gradient expansion far from the nodes, where relativistic invariance is lost.

One may conclude that the same UV-IR correspondence takes place for the finite temperature Nieh-Yan-anomaly. In this case, the temperature $T^2$ is an IR energy scale at which the quasi-relativistic fermions are well-defined, whereas $\Lambda^2$ is an explicit and non-universal UV cut-off \cite{Nissinen2019}. This correspondence can be explicitly verified from the hydrodynamic conservation laws, which exactly correspond to the Nieh-Yan anomaly, such as Eq. (\ref{dotP2}). The first attempt was made in Ref. \cite{NissinenVolovik2019}, where the finite temperature and Fermi-liquid corrected hydrodynamic transport parameters calculated in Ref. \cite{Cross1975} have been used.

\section*{Acknowledgements}
This work has been supported by the European
Research  Council  (ERC)  under  the  European  Union’s
Horizon 2020 research and innovation programme (Grant
Agreement No.  694248).       


\end{document}